\documentclass[twocolumn,prl,showpacs]{revtex4-1}
\usepackage[latin9]{inputenc}
\usepackage{amsmath}
\usepackage{amssymb}
\usepackage{graphicx}

\makeatletter

\@ifundefined{textcolor}{}
{%
 \definecolor{BLACK}{gray}{0}
 \definecolor{WHITE}{gray}{1}
 \definecolor{RED}{rgb}{1,0,0}
 \definecolor{GREEN}{rgb}{0,1,0}
 \definecolor{BLUE}{rgb}{0,0,1}
 \definecolor{CYAN}{cmyk}{1,0,0,0}
 \definecolor{MAGENTA}{cmyk}{0,1,0,0}
 \definecolor{YELLOW}{cmyk}{0,0,1,0}
}

\usepackage{amsfonts}
\usepackage{epsfig}
\usepackage{float}
\usepackage{latexsym}
\usepackage{graphicx}
\usepackage{color}

\makeatother

\begin{document}

\title{Where are the roots of the Bethe Ansatz equations?}

\author{R. S. Vieira}

\author{A. Lima-Santos}

\affiliation{Universidade Federal de São Carlos, Departamento de Física, CEP 13560-905,
São Carlos, Brazil}

\date{\today }
\begin{abstract}
Changing the variables in the Bethe Ansatz Equations (\textsc{bae})
for the \textsc{xxz} six-vertex model we had obtained a coupled system
of polynomial equations. This provided a direct link between the\textsc{
bae} deduced from the Algebraic Bethe Ansatz (\textsc{aba}) and the
\textsc{bae} arising from the Coordinate Bethe Ansatz (\textsc{cba}).
For two magnon states this polynomial system could be decoupled and
the solutions given in terms of the roots of some self-inversive polynomials.
From theorems concerning the distribution of the roots of self-inversive
polynomials we made a thorough analysis of the two magnon states,
which allowed us to find the location and multiplicity of the Bethe
roots in the complex plane, to discuss the completeness and singularities
of Bethe's equations, the ill-founded string-hypothesis concerning
the location of their roots, as well as to find an interesting connection
between the\textsc{ bae} with Salem\textquoteright s polynomials.
\end{abstract}


\maketitle
Exactly integrable models provide benchmarks for different areas of
physics as statistical mechanics \cite{baxter1}, condensed matter
physics \cite{essler}, quantum field theory \cite{abdalla}, nuclear
physics \cite{nuclear}, atomic-molecular physics \cite{angela} and
more recently for high energy physics through the gauge theory, string
theory and super-Yang-Mills theories \cite{zarembo}. An important
tool is the algebraic Bethe Ansatz \cite{takhfadd} culminating in
the Bethe Ansatz equations \cite{bethe,lieb}. Yet analytic results
have been unable to scale the unsurmountable wall of find their roots:
these have treated mostly by numerical methods, which are in general
hard to implement.

Here we analyze the solutions of the Bethe equations using some theorems
regarding self-inversive polynomials in order to answer the question
made in the title. 

The \textsc{ba} equations for the xxz six-vertex model on a $L\times L$
lattice, as deduced from the \textsc{aba} are 
\begin{eqnarray}
\left(\frac{\sinh(\lambda_{i}+\eta)}{\sinh\lambda_{i}}\right)^{L} & = & \prod_{k\neq i}^{N}\frac{\sinh(\lambda_{i}-\lambda_{k}+\eta)}{\sinh(\lambda_{i}-\lambda_{k}-\eta)},\nonumber \\
i & = & 1,2,...,N.\label{bae.1}
\end{eqnarray}
The solutions $\left\{ \lambda_{1},\lambda_{2},...,\lambda_{N}\right\} $
of (\ref{bae.1}) will furnish all $2^{L}$ states of the transfer
matrix for a lattice of $L$ columns.

Multiplying side-by-side these equations we get 
\begin{equation}
\prod_{i=1}^{N}\left(\frac{\sinh(\lambda_{i}+\eta)}{\sinh\lambda_{i}}\right)^{L}=1,\label{bae.2}
\end{equation}
 which suggests the following changing of variables, 
\begin{equation}
\frac{\sinh(\lambda_{i}+\eta)}{\sinh\lambda_{i}}=c_{i},\quad i=1,2,...,N,\label{bae.3}
\end{equation}
so that the $c_{i}$ should be subject to the constraint equation
\begin{equation}
c_{1}^{L}c_{2}^{L}\cdots c_{N}^{L}=1,\label{vinc}
\end{equation}
 which reflects the translational invariance of the periodic lattice.

Now, Eq. (\ref{bae.3}) can be easily solved for the rapidities $\lambda_{i}$,
\begin{equation}
\lambda_{i}=\mbox{arctanh}\left(\frac{\sinh\eta}{c_{i}-\cosh\eta}\right)=\frac{1}{2}\ln\left(\frac{c_{i}-\mathrm{e}^{-\eta}}{c_{i}-\mathrm{e}^{\eta}}\right),\label{Lambda}
\end{equation}
where the functions $\mbox{arctanh}(z)$ and $\ln(z)$ must be regarded
as multivalued complex functions, each branch differing by multiples
of $i\pi$. The new variables $c_{i}$ should yet to be determined.
To this end we insert (\ref{Lambda}) back on (\ref{bae.1}), which
gives us a system of $N$ polynomial equations, 
\begin{eqnarray}
c_{i}^{L} & = & (-1)^{N-1}\prod_{k\neq i}^{N}\frac{c_{i}c_{k}-2\Delta c_{i}+1}{c_{i}c_{k}-2\Delta c_{k}+1},\nonumber \\
\Delta & = & \cosh\eta,\quad i=1,2,..,N.\label{bae.5}
\end{eqnarray}

Here we observe that writing $c_{i}=\exp(k_{i})$, where $k_{i}$
are the Bethe's momenta, leaves (\ref{bae.5}) exactly equal to the
\textsc{bae} derived in the \textsc{cba} for the{\small{} }\textsc{\small{}xxz}
six-vertex model \cite{lieb}. Therefore, the relations (\ref{Lambda})
establish a direct link between the Bethe states of the \textsc{aba}
and the Bethe wave-functions of the \textsc{cba}. Thus, all the results
about completeness, singularities etc. which are valid for \textsc{cba,}
as obtained by Baxter in \cite{baxter}, will be also valid for the
algebraic version.

For $N=1$ the equations (\ref{bae.5}) reduce to $c_{1}^{L}=1$ and
$c_{1}$ is one of the $L$ roots of unity. This means that in a periodic
lattice of $L$ columns, the free pseudo-particle (magnon) has $L$
different rapidities given by the Bethe roots (\ref{Lambda}).

For $N=2$ we have three coupled equations for $c_{1}$ and $c_{2}$,
\begin{eqnarray}
c_{1}^{L}\left(c_{1}c_{2}-2\Delta c_{2}+1\right)+c_{1}c_{2}-2\Delta c_{1}+1 & = & 0,\nonumber \\
c_{2}^{L}\left(c_{1}c_{2}-2\Delta c_{1}+1\right)+c_{1}c_{2}-2\Delta c_{2}+1 & = & 0,\nonumber \\
c_{1}^{L}c_{2}^{L} & = & 1.\label{bae.7}
\end{eqnarray}
From the constraint equation we can set $c_{2}=\omega_{a}/c_{1}$
and the system becomes reduced to 
\begin{equation}
P_{a}(c_{1})=0,\qquad P_{a}(c_{1})/c_{1}^{L}=0,\label{S2}
\end{equation}
where $P_{a}(c_{1})$ are the following $L$-degree polynomials, 
\begin{eqnarray}
P_{a}(c_{1}) & = & (1+\omega_{a})c_{1}^{L}-2\Delta\omega_{a}c_{1}^{L-1}-2\Delta c_{1}+(1+\omega_{a}),\nonumber \\
\omega_{a} & = & \mathrm{e}^{\frac{2i\pi}{L}a},\qquad a=1,2,...,L.\label{PA}
\end{eqnarray}
Notice that $P_{a}(c_{1})$ satisfies $P_{a}(c_{1})=c_{1}^{L}P_{a}(\omega_{a}/c_{1})$,
the pair $\left(c_{1},c_{2}\right)$ and $\left(c_{2},c_{1}\right)$
therefore representing the same solution of (\ref{bae.7}). Moreover
$P_{a}(c_{1})$ is a self-inversive polynomial since they satisfy
$P_{a}(c_{1})=\omega_{a}c_{1}^{L}\overline{P}_{a}(1/c_{1})$, where
the bar means complex-conjugation. 

It seems that explicit solutions (in terms of radicals) of the equation
$P_{a}(c_{1})=0$ can be written only for small values of $L$, or
for very special values of $a$ and $\Delta$. The roots of $P_{a}(c_{1})$
of course can be easily obtained numerically. It is the self-inversive
property of the polynomials $P_{a}(c_{1})$ that provides crucial
informations about the location of the Bethe roots. 

We know from the theory of self-inversive polynomials \cite{marden}
that their roots are all symmetric with respect to the complex unit
circle $U=\{z\in C:|z|=1\}$. For the polynomials $P_{a}(c_{1})$,
however, we get a better scenario, since their roots can be distributed
only into two ways, namely, either they are all on $U$ or only two
dual roots $s$ and $\omega_{a}/s$ are not on $U$. In fact, the
exact behavior of the roots depends on the values of $\Delta$ from
which two are critical, 
\begin{equation}
\Delta_{1}=\left|\frac{1+\omega_{a}}{2}\right|\quad\mathrm{and}\quad\Delta_{2}=\frac{L}{L-2}\left|\frac{1+\omega_{a}}{2}\right|.\label{bae.9}
\end{equation}
 and this behavior can be summarized as follows,

\emph{i}) If $|\Delta|<\Delta_{1}$ then all roots of $P_{a}(c_{1})$
are on $U$ and they are simple;

\emph{ii}) If $|\Delta|>\Delta_{2}$ then only two dual roots of $P_{a}(c_{1})$
are out $U$ and all roots are simple;

\emph{iii}) If $|\Delta|=\Delta_{1}$ then all roots of $P_{a}(c_{1})$
are on $U$ but two roots on $U$ can be coincident;

\emph{iv}) If $|\Delta|=\Delta_{2}$ then two roots of $P_{a}(c_{1})$
are out $U$ but three roots on $U$ can be coincident;

\emph{v}) If $\Delta_{1}<|\Delta|<\Delta_{2}$ The roots are distributed
as in \emph{i}) or \emph{ii}), depending on the values of $L$ and
$a$.

The proof of the statements \textit{i}) and \textit{ii}) can be obtained
from theorems presented by Vieira in \cite{ricardo}, which are generalizations
of theorems presented previously by Lakatos and Losonczi in \cite{poly1};
\emph{iii}) and \emph{iv}) can be verified directly from the factorization
of $P_{a}(c_{1})$ and \emph{v}) was verified numerically only (since
the theorems mentioned above does not hold in this case). The exact
behavior in the case \emph{v}) is the following: all roots of $P_{a}(c_{1})$
will be in $U$ if, for even $L$, $a$ is odd and for odd $L$, if
$a$ is odd and $a<(L-1)/2$ or when $a$ is even and $a>(L+1)/2$;
for $a=(L-1)/2$ or $a=(L+1)/2$ the roots are all in $U$ if $L\equiv3\,(\text{mod }4)$;
otherwise the roots are in $U$, except for two dual roots $s$ and
$\omega_{a}/s$.

\begin{figure}
\includegraphics[scale=0.5]{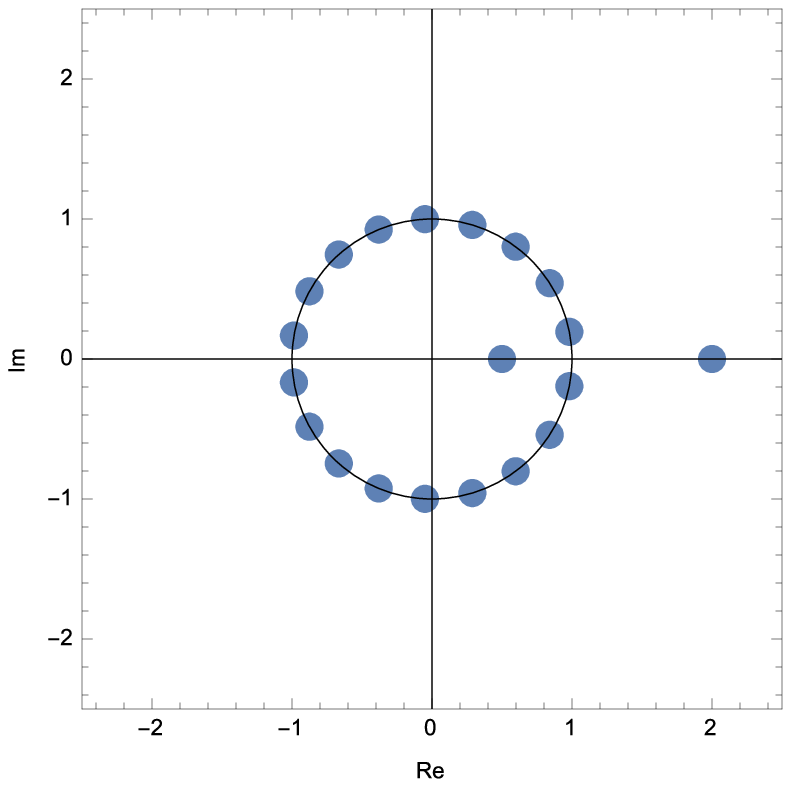}\includegraphics[scale=0.5]{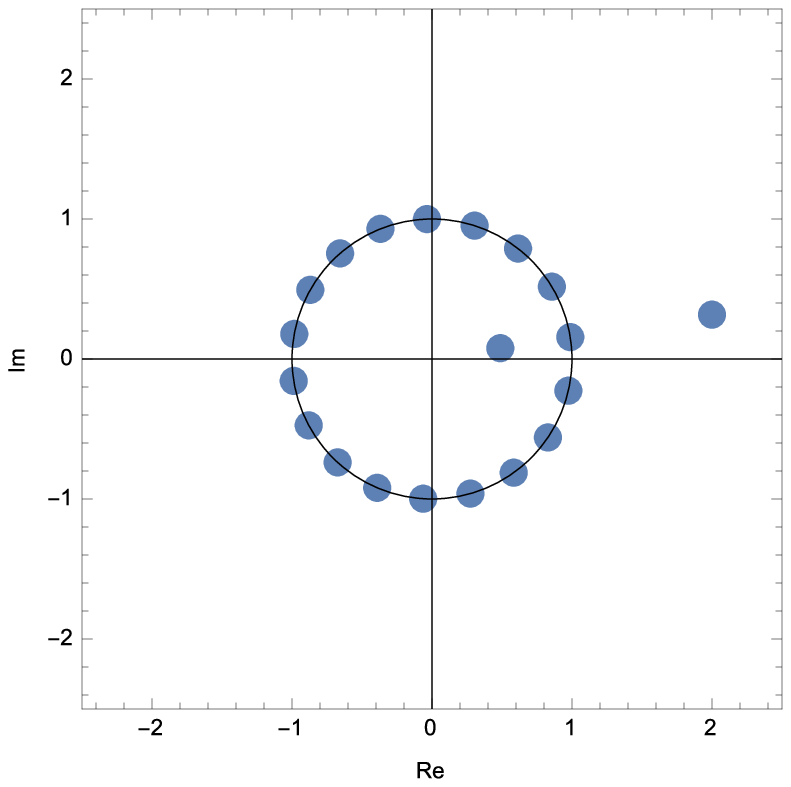} 

\protect\caption{Roots of $P_{L}(c_{1})$ and $P_{1}(c_{1})$ for $L=20$ and $\Delta=2$.
The first figure shows that $P_{L}(c_{1})$ is a Salem's polynomial.
In the second figure we have $P_{1}(c_{1})$ behaves as a ``rotated''
Salem's polynomial.\label{fig:Salem}}
\end{figure}

\begin{figure}
\includegraphics[scale=0.63]{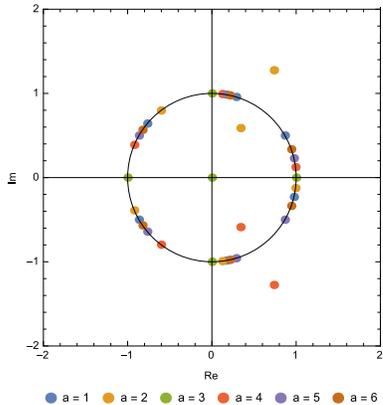}

\protect\caption{The roots of $P_{a}(c_{1})$ for $L=6$ and $\Delta=2/3$. For $a=\{1,5,6\}$
we have $|\Delta|<\Delta_{1}$, so the roots are all in $U$. For
$a=\{2,4\}$ we have $\Delta_{1}<|\Delta|<\Delta_{2}$ and since $a$
is even we have in each case two dual roots out of $U$. For $a=3$
we have $|\Delta|>\Delta_{2}$, then two roots are also out of $U$
(these roots are $0$, and $-\infty$). \label{fig:P6}}
\end{figure}

\begin{figure}
\includegraphics[scale=0.63]{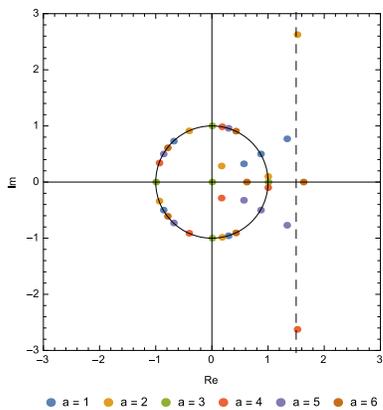}

\protect\caption{The roots of $P_{a}(c_{1})$ for $L=6$ and $\Delta=3/2$. For all
values of $a$ we have $|\Delta|\geq\Delta_{2}$ (the equality occurs
only for $P_{6}(c_{1})$, which does not have three roots coincident).
Therefore all $P_{a}(c_{1})$ have two dual roots out of $U$. \label{fig:P6-1}}
\end{figure}

Notice that in the case in the case \emph{i}), where the roots of
$P_{a}(c_{1})$ are all in $U$, if the coefficients of $P_{a}(c_{1})$
are integers then follows from a celebrated theorem of Kronecker \cite{Smith}
that all roots of $P_{a}(c_{1})$ are indeed roots of unit and therefore
they can be expressed by radicals. Moreover, in\emph{ ii}) the polynomials
$P_{a}(c_{1})$ have only two dual roots out of $U$. In algebraic
number theory a polynomial (with integer coefficients) whose roots
are all on the complex unit circle except for two positive reciprocal
roots $r$ and $1/r$ is named a Salem polynomial. Therefore the polynomials
$P_{L}(c_{1})$ are Salem's polynomials when $\Delta$ is a positive
integer greater than $1$. See Fig. \ref{fig:Salem}, \ref{fig:P6}
and \ref{fig:P6-1}. This is an interesting and non expected relation
between the \textsc{bae} for the two magnon state and Salem's polynomials,
since they are found only in a few fields of mathematical physics,
for instance, in Coxeter systems and the $(-2,3,7)$-pretzel knot
theory \cite{salem}. 

\begin{figure}
\includegraphics[scale=0.55]{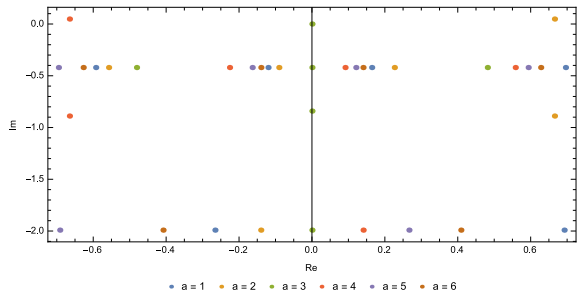}

\protect\caption{Bethe's Roots for $L=6$ and $\Delta=2/3$. Notice that the roots
of $P_{a}(c_{1})$ which lie on $U$ are mapped on the lines $y=-|\eta|/2$
and $y=-|\eta+i\pi|/2$. This leads to 1-string solutions. The dual
roots of $P_{a}(c_{1})$ which are out of $U$ are led to pairs with
the same real part, so they are related to 2-string solutions. \label{fig:L6}}
\end{figure}

\begin{figure}
\includegraphics[scale=0.55]{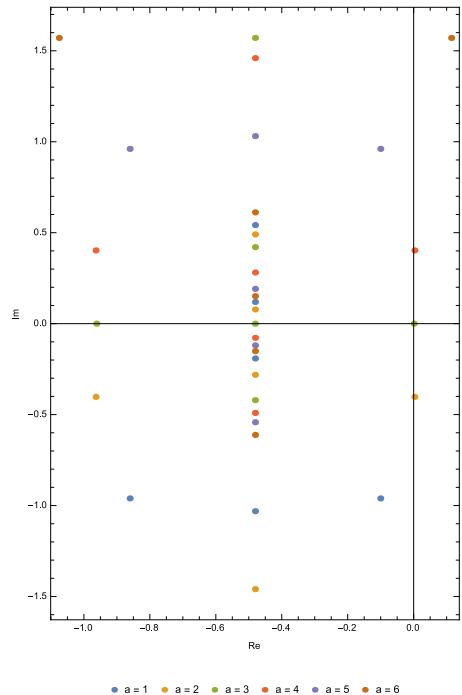}

\protect\caption{Bethe's Roots for $L=6$ and $\Delta=3/2$. Now the roots of $P_{a}(c_{1})$
which lie on $U$ are mapped on the line $x=-|\eta|/2$. This leads
to 1-string solutions. The dual roots of $P_{a}(c_{1})$ which are
out of $U$ are led to pairs with the same imaginary parts. This leads
to 2-string solutions.\label{fig:L6-1}}
\end{figure}

The analysis above gives us the distribution of the roots of $P_{a}(c_{1})$.
The correspondent location of the Bethe roots $\lambda_{i}$ is provided
by the formula (\ref{Lambda}), which can be seen as a conformal mapping
from the complex variables $c_{i}$ to $\lambda_{i}$. In fact, if
$|\Delta|>1$, so that $\eta$ is real, this mapping will send the
complex unit circle $U$ into the vertical line $x=-|\eta|/2$. On
the other hand, if $|\Delta|<1$ then (\ref{Lambda}) will map $U$
into a horizontal lines $y=-|\eta|/2$ and $y=-|\eta+i\pi|/2$. Notice
that usually the Bethe roots are expected to be arranged (in the thermodynamical
limit $L\gg1$) into groups of the same real part called strings \cite{viola1}.
A $n$-length string is a group of $n$ Bethe roots all of them with
the same real part. From what was said above we can see that, for
$|\Delta|<1$, the roots of $P_{a}(c_{1})$ which lie on $U$ will
lead to 1-string solutions, while the two dual roots outside $U$
will lead to 2-string solutions. The same will be true for the case
$|\Delta|>1$ provided we redefine a $n$-length string as group of
$n$ Bethe roots all of them with the same imaginary part. See Fig.
\ref{fig:L6} and \ref{fig:L6-1}. Notice as well that several violations
of the string hypothesis were already reported \cite{viola1}. This
not happens in our approach, since the string hypothesis is not used
in our deduction.

Here we remark the importance of this analysis for the range of $\Delta$
in the study of the completeness of the Bethe states and in the string
hypothesis. 

Let us first to consider $|\Delta|\neq\Delta_{1}$ and $|\Delta|\neq\Delta_{2}$.
For odd $L$ we have that $P_{a}(c_{1})$ factors to $P_{a}(c_{1})=(c_{1}+\omega_{b})Q_{a}(c_{1})$,
where $Q_{a}(c_{1})$ are $L$ self-inversive polynomials of $L-1$
degree and $2b\equiv a\text{ (mod }L)$. However, the solutions $c_{1}=-\omega_{b}$
lead us to $\lambda_{1}=\lambda_{2}$, that is, to $L$ states not
belonging to the two magnon sector. Hence, the wanted solutions are
the $L-1$ roots of each polynomial $Q_{a}(c_{1})$. A half of these
solutions are related by permutations, so we get the exact number
of solutions for the two magnon sector for odd $L$, namely, $L(L-1)/2$.

This same number of physical states for $L$ even $(L=2k)$ is obtained
by a more elaborate sum of terms. This happens because the case $\omega_{a}=-1$
is special, since $P_{a}(c_{1})$ collapses in this case to 
\begin{equation}
P_{L/2}(c_{1})=2\Delta c_{1}(c_{1}^{L-2}-1).
\end{equation}
The possibility $c_{1}=0$ leads to $c_{2}=-\infty$ and the system
(\ref{S2}) is not actually satisfied. However, this would lead to
the finite Bethe roots $\left(\lambda_{1},\lambda_{2}\right)=\left(-\eta,0\right)$
which is in fact a solution of the Bethe equations. The explanation
is that we had implicitly assumed $c_{j}\neq0$, $1\leq j\leq L$,
in deriving (\ref{bae.3}) and thus these cases must be analyzed separately.
Thanks to the conjugation property, we have as well the solution $\left(\lambda_{1},\lambda_{2}\right)=\left(0,-\eta\right)$,
so we get two additional solutions of the \textsc{bae} which must
be taken into account. 

For the other case, $c_{1}$ must be a $(L-2)$-root of unity and
we get 
\begin{equation}
\lambda_{1}=\frac{1}{2}\ln\left(\frac{\zeta_{j}-\mathrm{e}^{-\eta}}{\zeta_{j}-\mathrm{e}^{\eta}}\right),\quad\lambda_{2}=\frac{1}{2}\ln\left(\frac{\zeta_{j}\mathrm{e}^{-\eta}+1}{\zeta_{j}\mathrm{e}^{\eta}+1}\right),
\end{equation}
where $\zeta_{j}=e^{\frac{2\pi i}{L-2}j}$, for $j=1,2,...L-2$. In
this case we have $\lambda_{2}\neq\lambda_{1}$ except when $j=(k-1)/2$
or $j=3(k-1)/2$, which can only happen if $k$ is odd. Therefore,
the case $a=k$ furnishes $2(k-1)$ solutions if $k$ is even, but
only $2(k-2)$ solutions if $k$ is odd. Now let us consider the cases
where $a\neq k$. Here follows that the equations (\ref{PA}) factor
to $P_{a}(c_{1})=(c_{1}^{2}-\omega_{a})Q_{a}(c_{1})$ when $a$ is
odd, but it not factor when $k$ is even. For even $k$ we have $k-1$
cases where $a$ is even and $k$ cases where $a$ is odd, which gives
us $(k-1)\left(2k\right)+\left(k\right)\left(2k-2\right)$ solutions.
For odd $k$ we have $k$ cases where $a$ is even and $k-1$ cases
where $a$ is odd, which gives us $(k)\left(2k\right)+(k-1)(2k-2)$
solutions. Taking into account the solutions from $a=k$ and that
half of the solutions are related by permutations we get, in both
cases, a total of $k(2k-1)=L(L-1)/2$ solutions, as expected. 

Here we remark that for $L=2$, $P_{1}(c_{1})$ is identically satisfied
and we have a physical state with a free parameter (see \cite{baxter}
for more details). Moreover, it turns out that $P_{2}(c_{1})$ has
only the singular solutions $c_{1}=e^{\pm\eta}$, as the solutions
studied in \cite{nepo}.

Now let us consider the cases where multiple roots appears. If $|\Delta|=\Delta_{1}$,
here we have that, for $L=2k-1$, $(k>1$), there are $k-1$ polynomials
($a=2,4,...,k-2,k+1,k+3,...,L$) for $k$ even and $k$ polynomials
($a=2,4,...,k-1,k,k+2,k+4,...,L$) for $k$ odd which factor in the
form $P_{a}(c_{1})=(1+\omega_{a})(c_{1}+\omega_{b})(c_{1}-\omega_{b})^{2}Q_{a}(c_{1})$,
with $2b\equiv a\text{ (mod }L)$. For $L=2k$, $(k>1)$, $P_{a}(c_{1})$
factors as before but without double roots if $a$ is odd. It turns
out that for $a$ even it now factors to $P_{a}(c_{1})=(1+\omega_{a})(c_{1}-\omega_{f})^{2}R_{a}(c_{1})$
and we have $k-1$ polynomials of this type if $k$ is even and $k$
polynomials if $k$ is odd, all with two coincident roots identified
by the indices $a=2p$ and $f=p$ for $a<k$ and $f=k+p$ for $a>k$.
Summarizing, the number of states for two magnon in a lattice of $L$
columns is reduced to $L(L-1)/2-\alpha$ where $\alpha=k-1$ for $k$
even and $\alpha=k$ for $k$ odd when $L=2k-1$ and $\alpha=2k-1$
when $L=2k$. A similar count holds when $|\Delta|=\Delta_{2}$. In
this case however the number of states is reduced to $L(L-1)/2-\beta$
where $\beta=k$ for $k$ even and $\beta=k-1$ for $k$ odd when
$L=2k-1$ and $\beta=2k$ for $k$ even and $\beta=2(k-1)$ for $k$
odd when $L=2k$.

The presence of these multiple roots in the two magnon sector mask
the completeness \cite{viola1}. However, we can use the $S$-matrix
language in order to understand these new states not as two magnon
states, but as free bound states of two pseudo-particles with the
same or parallel rapidities ($c_{1}^{L}=c_{2}^{L}=1$). This is another
physical problem.

Finally, a short comment about the case $N=3$. Now we have three
coupled equations in $c_{1}$, $c_{2}$ and $c_{3}$ plus the constraint
equation $c_{1}^{L}c_{2}^{L}c_{3}^{L}=1$. In particular, self-inversive
polynomials with $\omega_{a}=\varepsilon=\pm1$ can be obtained by
setting $c_{3}=1$ and $c_{2}=\varepsilon/c_{1}$ and it has the form
\begin{eqnarray}
P_{L}(c_{1}) & = & c_{1}^{L+1}-\left(3\varepsilon\Delta-1\right)c_{1}^{L}+\varepsilon\Delta\left(2\varepsilon\Delta-1\right)c_{1}^{L-1}\nonumber \\
 & - & \Delta\left(2\varepsilon\Delta-1\right)c_{1}^{2}+\left(3\varepsilon\Delta-1\right)c_{1}-\varepsilon=0.
\end{eqnarray}
In fact, we have verified that all roots of $P_{L}(c_{1})$ are in
$U$ if $0\leq\Delta\leq1$, but for other intervals we found that
$P_{L}(c_{1})$ may have two or four roots out of $U$. Moreover,
by removing multiple roots of these polynomials (for each value of
$\varepsilon$) we got $(L-1)(L-2)/6$ states for $L\geq6$ and taking
into account the other values of $\omega_{a}$ the total is $L(L-1)(L-2)/6$
states. We will present a more detailed study of the $N\geq3$ case
in a forthcoming study. This study can be also generalized to more
general vertex models as the eight-vertex model and those requiring
the nested \textsc{bae}.


(\textit{Acknowledgments}). It is ALS's pleasure to thank professor
Roland Köberle for their help and advice in preparing this article.
The work of RSV has been supported by São Paulo Research Foundation
(FAPESP), grant \#2012/02144-7. ALS also thanks Brazilian Research
Council (CNPq), grant \#310625/ 2013-0 and FAPESP, grant \#2011/18729-1
for financial support.


\end{document}